\documentclass{article}

\usepackage[final]{nips_2018}

\usepackage[utf8]{inputenc} 
\usepackage[T1]{fontenc}    
\usepackage{hyperref}       
\usepackage{url}            
\usepackage{booktabs}       
\usepackage{amsfonts}       
\usepackage{nicefrac}       
\usepackage{microtype}      
\usepackage{graphicx}

\title{Neural Wavetable: a playable wavetable synthesizer using neural networks}

\author{
  Lamtharn Hantrakul\thanks{Currently an AI Resident with Google Brain} \quad Li-Chia Yang \\
  Center for Music Technology\\
  Georgia Institute of Technology\\
  Georgia, USA 30318 \\
  \texttt{[lhantrakul3][richard40148]@gatech.edu} \\
}

\begin{document}

\maketitle

\begin{abstract}
    We present Neural Wavetable, a proof-of-concept wavetable synthesizer that uses neural networks to generate playable wavetables. The system can produce new, distinct waveforms through the interpolation of traditional wavetables in an auto-encoder's latent space. It is available as a VST/AU plugin for use in a Digital Audio Workstation.
\end{abstract}

\section{Introduction}

WaveNet, a deep neural network architecture for speech and audio generation, has brought exciting innovations in the sound domain since its introduction in 2016 [2]. In the musical domain, WaveNet has been used as an auto-encoder capable of interpolating -- or non-linear ``blending'' -- different instrument sounds through NSynth [1].

Inspired by this approach, we present a proof-of-concept system addressing a smaller scale audio task: generating short wavetables. Rather than generate long segments of speech and conditioning the network on long-term periodicity, we focus on generating short wavetables of length of 512 samples; a contrast to the tens of thousands of samples required even in just a second of audio. These short wavetables are fed into a real-time wavetable synthesizer implemented in C++ and the JUCE framework. The source-code and VST/AU plug-in are available for public download\footnote{Neural Wavetable. \url{https://github.com/RichardYang40148/Neural_Wavetable_Synthesizer}}.

\section{Background}
Wavetable synthesis is an industry-standard approach which uses a bank of ``wavetables'' of lengths between 512-2048 and sample-rate conversion techniques to efficiently generate musical sounds. In context of this work, wavetable synthesis efficiently handles real-time generation of different pitches given a single wavetable generated by a neural network. For brevity, readers should reference the original works for a complete description of Nsynth [1], WaveNet [2] and Wavetable synthesis [3].

\begin{figure}
  \centering
  \includegraphics[width=\linewidth]{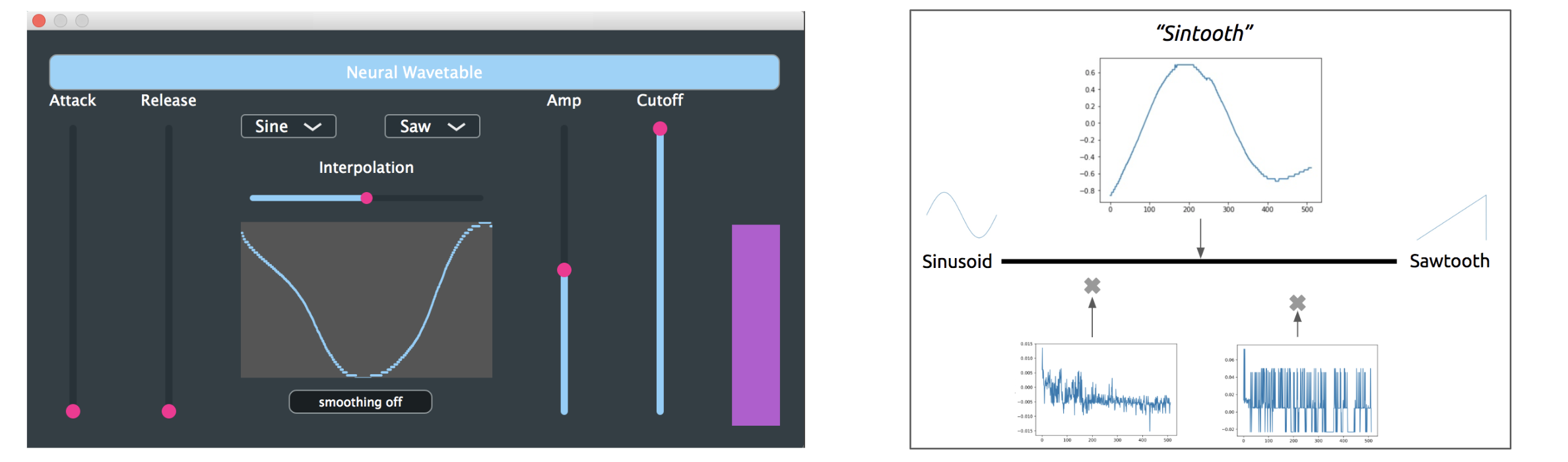}
  \caption{\textbf{Left:} Neural Wavetable plug-in GUI with interpolation, envelope and filter controls. \textbf{Right:} Interpolated vectors along an edge connecting two embedding vectors produces an interpolated wavetable that is musically playable. Deviating off this edge produces noise.}
\end{figure}

\section{Implementation Details}
\subsection{WaveNet as a wavetable autoencoder}

In this proof-of-concept, we use the pre-trained model NSynth WaveNet model as-is [1]. Given the task of generating a short wavetable is  smaller than NSynth's original task of generating several seconds of audio, we thus first perform qualitative tests exploring the latent space using wavetables instead of audio segments (Figure 1).

First, we found the model retains short-term coherence: it can decode playable wavetables from inputs of short, simple wavetables such as sinusoid, triangle and sawtooth waves of length 512 (Figure 1). The model also produces a reconstruction 180$^\circ$ out of phase from the input. This is a peculiar finding, given how the pre-trained model is trained on raw audio and not the fourier representation, where it is normal to disregard phase. Either way, this does not affect perception of pitch and timbre, since the human auditory system is largely phase-independent [4].

Secondly, interpolating along the edge connecting two embedding vectors from a sinusoid and sawtooth yielded waveforms that were a mix of these two input wavetables. However, moving ``off'' this edge produced noise or quantization noise.

\subsection{Real-time limitations}
The limitation of real-time rendering occurs in the encoding and decoding process of the pre-trained model (approximately 4 seconds to encode, and 7 seconds to decode a waveform on a GPU machine.). As a result, we pre-rendered 300 wavetables interpolating between three combinations of sinusoid, saw and triangle waveforms, and perform the wavetable synthesis in real-time.
This approach is similar to the pre-generated waveforms in the public-facing NSynth demo\footnote{NSynth Sound Maker. \url{https://experiments.withgoogle.com/ai/sound-maker/view/}}.

\subsection{Wavetable synthesis engine}

Our real-time wavetable synthesizer uses bi-linear interpolation for reading wavetables. Since the generated wavetables vary in amplitude and offsets, we remove the mean and normalize to the range -1 to 1.  Each of the 512-sample wavetables are zero-padded on either side to produce a 514-sample wavetable. This ensures waveform looping starts and ends on the same value. In some wavetables, this introduces large discontinuities in the signal, leading to addition of high frequencies. We thus include an option to ``smooth'' the waveform at these discontinuities by approximating a line connecting extreme values to zero. The effect sounds like a gentle high-cut filter. The user can control these parameters via the GUI shown in Figure 1. 

\section{Conclusion and Future Work}

We encourage readers to experiment with the Neural Wavetable VST/AU plugin; one can generate interesting shades combining a bright sawtooth and subdued sinusoid. By leveraging the strengths of both traditional and new machine-learning enabled synthesis methods, a newer implementation of Neural Wavetable under development employs a lightweight model designed to interpolate and decode short wavetables in real-time using a variational autoencoder.

\small

[1] Engel, J., Resnick, C., Roberts, A. \ \& others \ (2017) Neural audio synthesis of musical notes with wavenet autoencoders. {arXiv preprint
arXiv:1704.01279}

[2] Oord, A., Dieleman, S. \ \& others \ (2016) Wavenet: A generative model for raw audio {\it arXiv preprint arXiv:1609.03499}

[3] Roads, C. \ \& Stawn, J. Barkai, E.\ (1996) {\it The Computer Music Tutorial} MIT Press.

[4] Zwicker, E. \ \& Fastl, H. (2013) {\it Psychoacoustics: Facts and Models} Vol. 22. Springer Science \& Business Media

\end{document}